%% file: PixeldEdx2016.tex
\pdfoutput=1
\newcommand*{\ATLASLATEXPATH}{}
\documentclass[texmf,cernpreprint,texlive=2011,txfonts,UKenglish]{\ATLASLATEXPATH atlasdoc}

\usepackage{\ATLASLATEXPATH atlaspackage}
\usepackage{\ATLASLATEXPATH atlasbiblatex}

\usepackage{\ATLASLATEXPATH atlascontribute}

\usepackage{\ATLASLATEXPATH atlasphysics}

\newcommand{\dedx}{\ensuremath{\mathrm{d}E/\mathrm{d}x}}
\newcommand{\dEdx}{\ensuremath{\mathrm{d}E/\mathrm{d}x}}
\newcommand{\lumi}{3.2}
\newcommand{\lumiUnc}{5\%}

\AtlasTitle{Search for metastable heavy charged particles
with large ionization energy loss in \emph{pp} collisions at $\sqrt{s} = 13$~\TeV\ using the \mbox{ATLAS} experiment}

\author{The ATLAS Collaboration}

\hypersetup{pdftitle={ATLAS document},pdfauthor={The ATLAS Collaboration}}

\AtlasRefCode{SUSY-2016-03}

\PreprintIdNumber{CERN-EP-2016-063}

 \AtlasJournal{Phys.\ Rev.\ D.}

\AtlasAbstract{%
This paper presents a search for massive charged long-lived particles produced in \emph{pp} collisions at $\sqrt{s}=13$~\TeV\ at the LHC using the \mbox{ATLAS} experiment. The dataset used corresponds to an integrated luminosity of $\lumi$~fb$^{-1}$.
Many extensions of the Standard Model predict the existence of massive charged long-lived particles, such as $R$-hadrons.
These massive particles are expected to be produced with a velocity significantly below the speed of light, and therefore to have a specific ionization higher than any Standard Model particle of unit charge at high momenta.
The Pixel subsystem of the \mbox{ATLAS} detector is used to measure the ionization energy loss of reconstructed charged particles and to search for such highly ionizing particles. 
The search presented here has much greater sensitivity than a similar search performed using the \mbox{ATLAS} detector in the $\sqrt{s}=$~8~\TeV\ dataset, 
thanks to the increase in expected signal cross-section due to the higher center-of-mass energy of collisions, 
to an upgraded detector with a new silicon layer close to the interaction point, 
and to analysis improvements.
No significant deviation from Standard Model background expectations
is observed, and lifetime-dependent upper limits on
$R$-hadron production cross-sections and masses are set. 
Gluino $R$-hadrons with lifetimes above 0.4~ns and decaying to $q\bar{q}$ plus a 100~\GeV\ neutralino are excluded at the $95\%$ 
confidence level, with lower mass limit ranging between 740~\GeV\ and 1590~\GeV. 
In the case of stable $R$-hadrons the lower mass limit at the $95\%$ confidence level is 1570~\GeV.
}

 \AtlasCoverAnalysisTeam{Bradley Axen, Dario Barberis, Andrea Favareto, Andrea Gaudiello, Claudia Gemme, Elisa Guido, Beate Heinemann, Laura Jeanty, Simone Pagan Griso, Stefano Passaggio, Leonardo Paolo Rossi}

 \AtlasCoverEdBoardMember{Markus~Elsing~(chair), Arely~Cortes-Gonzalez, Andy~Haas}

\begin{document}

\maketitle


\section{Introduction} \label{sec:Introduction}
Massive, long-lived particles (LLPs) are predicted by a wide range of physics models that extend the Standard Model (SM). LLPs arise in proposed solutions to the gauge hierarchy problem~\cite{Susskind:1982mw}, including supersymmetric (SUSY) models that either violate~\cite{Dreiner:1997uz,Berger:2003kc,Barbier:2004ez} or conserve~\cite{Fairbairn:2006gg,Kolda:1997wt,Baer:1998pg,Gates:1999ei,Mafi:1999dg,Chertok:2001ja,Giudice:2004tc,ArkaniHamed:2004yi} $R$-parity. The lifetime of these particles depends on the mass difference between the particle and the lightest stable SUSY particle, and on the size of any $R$-parity-violating coupling~\cite{Randall:2011}. 

Because of their large mass, LLPs are expected to be slow ($\beta$ significantly below 1) and, if charged, to have a specific ionization higher than any SM particles of unit charge at high momenta. The Pixel subsystem~\cite{Aad:2008zz} of the \mbox{ATLAS} detector~\cite{PERF-2007-01} provides measurements of ionization energy loss ($\dedx$) for charged particles and can be used to distinguish such highly ionizing particles from SM particles. In this paper, this information is used to search for LLPs using data collected at $\sqrt{s} = 13$~\TeV, extending in reach a study performed on data collected at $\sqrt{s} = 8$~\TeV~\cite{SUSY-2014-09}. The increase in probability for high-energy parton-parton collisions significantly enhances the production cross-section of LLPs via the strong interaction. Together with this enhancement, the search sensitivity has been considerably extended thanks to the upgrades to the Pixel detector (see Section~\ref{sec:Experiment}) and improvements to the analysis (see Section~\ref{sec:Selection}).

Many strategies have been explored to find LLPs at the LHC. A search using the complete $\sqrt{s} = 8$~\TeV\ dataset recorded by the \mbox{ATLAS} detector~\cite{SUSY-2013-22} combined Pixel $\dEdx$ information with time-of-flight measurements, made in both the calorimeters and the muon spectrometer, to search for LLPs that do not decay within the detector. Charginos that are nearly mass degenerate with the lightest stable particle have been searched for in ATLAS looking for disappearing tracks~\cite{SUSY-2013-01}. A displaced vertex search~\cite{SUSY-2014-02} performed by the ATLAS Collaboration provided limits on the mass of gluinos decaying due to couplings violating both $R$-parity and lepton number conservation. The CMS Collaboration published an analysis searching for nondecaying LLPs based on $\dedx$, $\beta$, and muon identification~\cite{CMS-EXO-11-022}. Additional analyses at the CMS experiment searched for LLPs with short lifetimes by looking for secondary vertices~\cite{CMS-EXO-12-038,CMS-EXO-12-037} and for disappearing tracks~\cite{CMS-EXO-12-034}.

The current study addresses many different models of new physics, in particular those that predict the production of metastable massive particles with \emph{O}(ns) lifetime at LHC energies, such as mini-split SUSY~\cite{Giudice:2004tc,ArkaniHamed:2004fb,Arvanitaki:2013} or anomaly-mediated supersymmetry-breaking (AMSB) models~\cite{ASMB1:Giudice1998,ASMB2:Randall1999}. A metastable gluino with a mass of approximately 1~\TeV\ would be compatible with the measured Higgs boson mass according to mini-split SUSY models, which also predict squarks with masses of $10^3$--$10^5$~\TeV, therefore making the gluino the only observable sparticle produced through the strong interaction at LHC energies. 
Results are presented assuming the production of $R$-hadrons as composite colorless states of a gluino together with SM quarks or gluons and the subsequent decay, on nanosecond timescales, of the gluino to a stable neutralino and a $q\bar{q}$ pair~\cite{Farrar:1978xj}.
This search could, in principle, also be sensitive to the production of LLPs through weak interactions, but these models are not considered here as the cross-sections are much lower.

In the rest of this paper, $R$-hadrons with lifetimes consistent with decaying inside the ATLAS detector are referred to as metastable; the search strategy focusses on decays occurring within the active detector volume, which covers lifetimes from around 1 ns to several tens of ns. $R$-hadrons whose lifetime disfavors decay within the ATLAS detector are termed stable. The approach described in this paper allows direct observation of charged $R$-hadrons which are either stable or metastable and traverse the first seven layers of the ATLAS inner tracking system.


%
\section{ATLAS Detector and Pixel \boldmath{$\dedx$} measurement}\label{sec:Experiment}
The ATLAS detector\footnote{ATLAS uses a right-handed coordinate system with its origin at the nominal interaction point in the centre of the detector and the $z$-axis coinciding with the axis of the beam pipe. The $x$-axis points from the interaction point to the centre of the LHC ring, and the $y$-axis points upward. Cylindrical coordinates ($r$, $\phi$) are used in the transverse plane, $\phi$ being the azimuthal angle around the $z$-axis. The pseudorapidity is defined in terms of the polar angle $\theta$ as $\eta$ = $-$ln tan($\theta$/2).} consists of a tracker, for measuring the trajectories of charged particles, surrounded by a solenoid magnet, followed by calorimeters for measuring the energy of particles that interact electromagnetically or hadronically. 
A muon spectrometer immersed in a toroidal magnetic field surrounds the calorimeters, and provides tracking for muons. The detector is hermetic and can therefore measure the missing transverse momentum associated with each event. A complete description of the ATLAS detector can be found in Ref.~\cite{PERF-2007-01}.
The tracker is made of three detector systems.  Moving from the solenoid magnet toward the beam collision region, there is first a transition radiation tracker with approximately 400,000 channels~\cite{ATLAS_TRT}, followed by a silicon microstrip detector (SCT) with about 6 million channels~\cite{IDET-2013-01},  and at the innermost radius sits a roughly 92 million channel silicon Pixel detector, which is crucial for this measurement, and is described below in some detail.

The current Pixel detector provides at least four precision measurements for each track in the region $|\eta| <2.5$ at radial distances of 3.4 to 13~cm from the LHC beam line. 
After Run 1, the ATLAS Pixel detector was upgraded  with the insertion of an additional layer, the Insertable B-Layer (IBL), which was installed inside the Run 1 Pixel detector, mounted on a new beam pipe of smaller diameter. The IBL has smaller area pixels, reduced thickness, and different coding electronics, which provides charge measurements with lower resolution and dynamic range than the other Pixel layers. It therefore requires a separate treatment of the $\dEdx$ information.
At normal incidence, the average charge released by a minimum ionizing particle (MIP) in a pixel sensor is $\approx$~20000~\textit{e}$^{-}$ ($\approx$~16000~\textit{e}$^{-}$ for IBL) and the charge threshold is set to 3500$\pm$40~\textit{e}$^{-}$ (2500$\pm$40~\textit{e}$^{-}$ for IBL). Signals are accepted if they are larger than this threshold. They are  then matched to a specific beam crossing. The hit efficiency under these conditions exceeds 99$\%$.
When detector data are read out, the Time over Threshold (ToT), i.e. 
the time interval with the signal above the threshold, is digitized with 8 bits (4 bits for IBL).
The ToT is approximately proportional to the ionization charge~\cite{BaBar_ToT} and its dynamic range corresponds to 8.5 times (1.5 times for IBL) the average charge released by a MIP for a track normal to the silicon detectors which deposits all its ionization charge in a single pixel. If this value is exceeded, the hit charge information is either underestimated in the IBL, where the electronics signals the excess with an overflow bit, or lost in the other pixel layers.

\begin{figure}
\resizebox{0.65\textwidth}{!}{%
\includegraphics{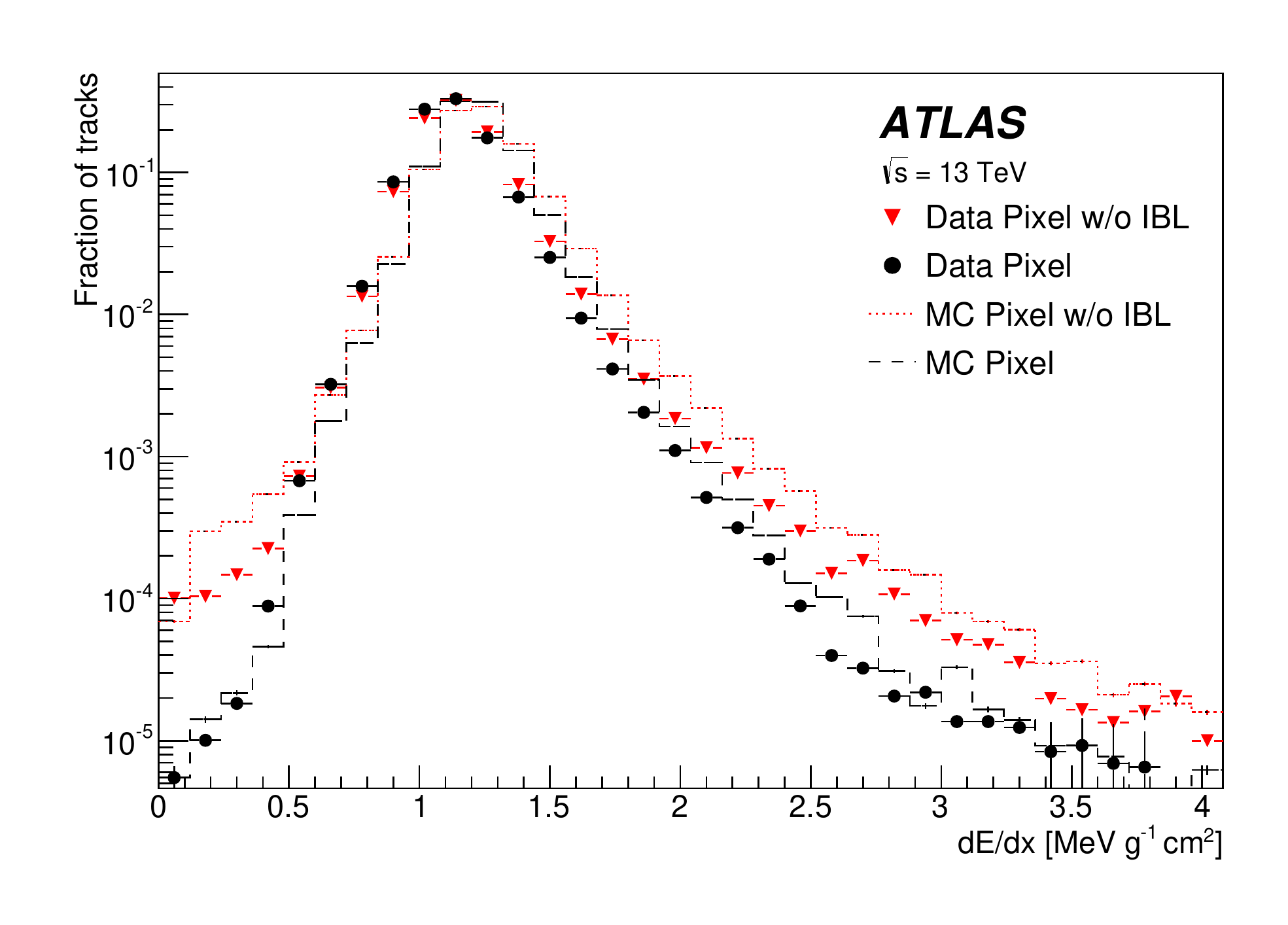}
}
\centering
\caption{Distribution of the $\dedx$ computed with the Pixel algorithm in data (black dots) and Pixel without IBL (red triangles) for minimum bias event tracks selected as in Ref.~\cite{Aad:2016mok}, with the additional requirements of $p > 3$~\GeV\ and at least two associated pixel clusters. The corresponding $\dedx$ distributions for simulation with and without the IBL are also plotted, respectively, as dashed and dotted lines. The simulation overestimates the $\dedx$, which is taken into account in the systematic uncertainty evaluation. The addition of IBL clusters reduces the tails at both low and high $\dedx$. The fraction of tracks with $\dedx<$0.5 ($>$1.8) \MeV\ g$^{-1}$ cm$^{2}$ decreases with the use of the IBL charge information from 0.065\% to 0.018\% (from 0.89\% to 0.44\%).}
\label{fig:includeIBL} 
\end{figure}

The charge released by a track crossing the Pixel detector is rarely contained within just one pixel; neighboring pixels registering hits are joined together using a connected component analysis~\cite{paperCCA,PERF-2012-05} to form clusters. The charge of a cluster is calculated by summing the charges of all pixels belonging to the cluster after calibration corrections; the IBL overflow information is accounted for separately. 
The $\dedx$ is estimated using the average of the individual cluster ionization measurements (charge collected in the cluster per unit track length in the sensor), for the clusters associated with a track. To reduce the effect of the tails of the Landau distribution, the average is evaluated after removing the highest $\dedx$ cluster, or the two highest $\dedx$ clusters in the rare case of more than four clusters on the track.
A track is considered to have a good ionization measurement only if there are at least two remaining clusters. Including the IBL charge information reduces the tails of the $\dEdx$ distribution, as illustrated in Figure~\ref{fig:includeIBL}, and increases the fraction of tracks for which the ionization measurement is valid from 77\% to 91\%.
The most probable value of the measured $\dedx$ is 1.12 \MeV\ g$^{-1}$ cm$^{2}$ for a minimum ionizing particle. 
This value is obtained with a gaussian fit to the data using the complete Pixel detector (see Figure~\ref{fig:includeIBL}); the variance of the gaussian fit is 0.13 \MeV\ g$^{-1}$ cm$^{2}$. 
The $\beta\gamma$ measurable with the current $\dEdx$ method cannot be smaller than approximately 0.3 for particles with unit charge because of the ToT dynamic range.

\section{Mass Calculation}\label{sec:Strategy}
The average energy loss of massive, charged particles in matter is expected to follow the Bethe--Bloch distribution, which can be expressed as a function of the $\beta\gamma$ of the LLPs.
Their mass can be derived from a fit of the measured specific energy loss and the reconstructed momentum to a parametric Bethe-Bloch distribution in the range $0.3< \beta\gamma<1.5$.
This range overlaps the expected average $\beta\gamma$ of LLPs produced at the LHC, which decreases with the particle mass from $\langle\beta\gamma\rangle\approx$~2.0 at 100~\GeV\ to  $\langle \beta\gamma\rangle \approx$~0.5 at 1600~\GeV. 
The parametric function describing the relationship between the most probable value of the energy loss $ (\dEdx)_\mathrm{MPV}$ and $\beta\gamma$ is:

\begin{equation}
(\dEdx)_\mathrm{MPV}(\beta \gamma) = {p_1\over \beta^{p_3}}\mathrm{ln}(1+[p_2\beta\gamma]^{p_5})-p_4 \label{bbfun}
\end{equation}
The \textit{p$_i$}, with $i=1\ldots5$, calibration constants were measured in this data set using low-momentum pions, kaons
and protons reconstructed in ATLAS as described in Ref.~\cite{ATLAS-CONF-2011-016}. 

\begin{figure}
\resizebox{0.65\textwidth}{!}{%
\includegraphics{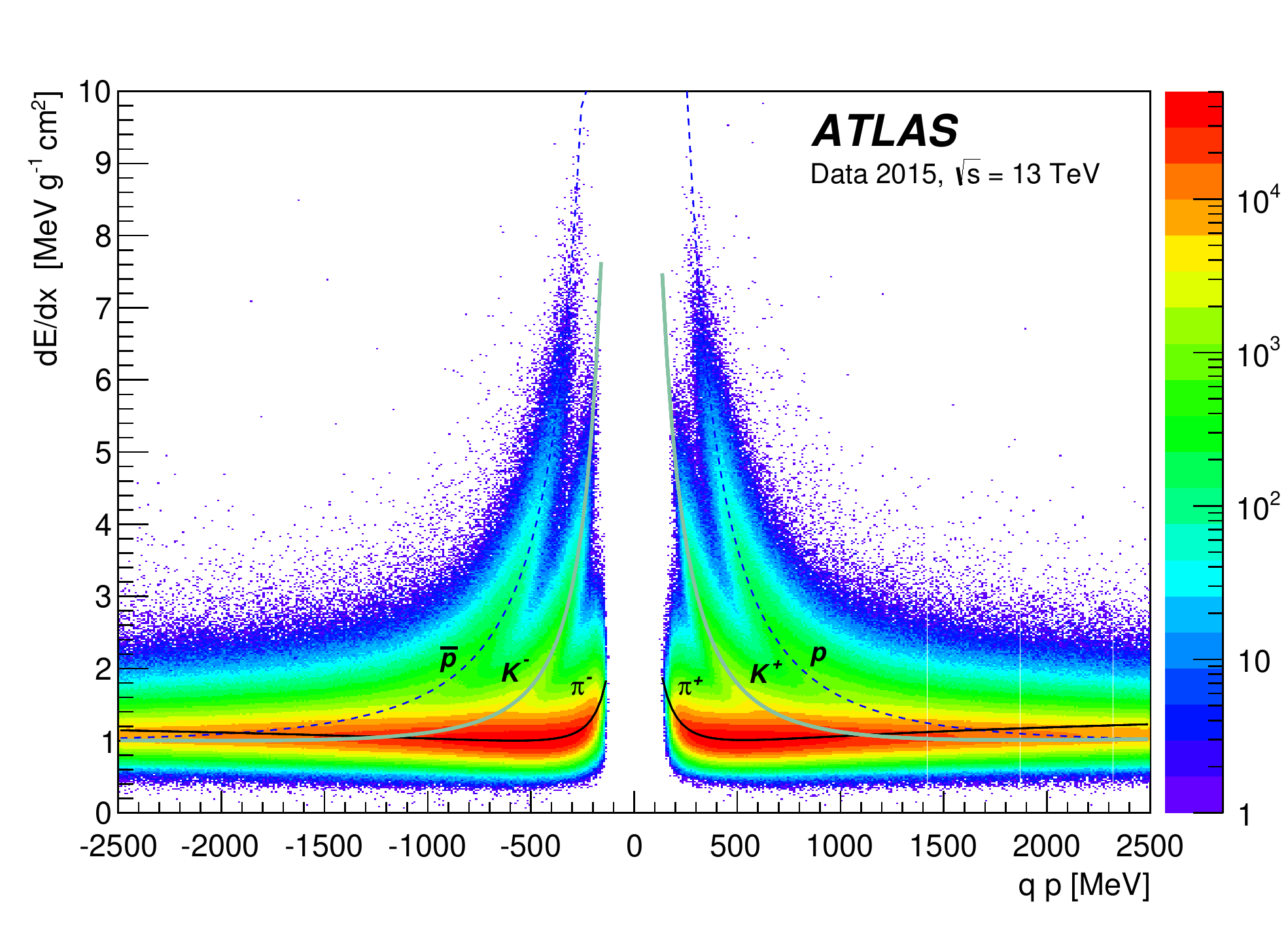}
}
\centering
\caption{Two-dimensional distribution of $\dedx$ versus charge signed momentum (qp) for minimum-bias event tracks compatible with a primary vertex, which has the same data selection as in Ref.~\cite{Aad:2016mok}. The distributions of the most probable value for the fitted probability density functions of pions, kaons and protons are superimposed.}
\label{fig:dEdxvsp}
\end{figure}

The distribution of the $\dedx$ versus momentum is shown in Figure~\ref{fig:dEdxvsp} together with the fits to the pion, kaon and proton masses.
The mass calculation performed through the $\dedx$  method is monitored by checking the stability of the proton mass measurement during the data taking.

A mass estimate $M$ is obtained by numerically solving
the equation~\ref{bbfun} 
for the unknown $M$. In simulated $R$-hadron events, the reconstructed mass is found to reproduce well the generated mass up to masses around 1800~\GeV, and a residual $3\%$ correction is applied to the reconstructed mass to improve the agreement. For particles with higher masses (beyond the current sensitivity of this analysis), the reconstructed mass is observed to slightly underestimate the generated mass.
The half-width at half maximum of the reconstructed mass distribution increases with the mass value. This is due to the momentum measurement uncertainty dominating the mass resolution for masses greater than $\approx$200~\GeV.


\section{$R$-hadron Simulation}\label{sec:Simulation}
A number of samples of simulated signal events are used in this analysis to determine the expected LLP signal efficiencies and to estimate uncertainties in the efficiency. 

For stable $R$-hadrons, pair production of gluinos with masses between 800~\GeV\ and 1800~\GeV\ is simulated in \textsc{Pythia 6.4.27}~\cite{Sjostrand:2006za} with the AUET2B~\cite{ATL-PHYS-PUB-2011-014} set of tuned parameters for the underlying event and the CTEQ6L1~\cite{ATLAS_PDF_set} parton distribution function (PDF) set, incorporating dedicated hadronization routines~\cite{Kraan:2004tz} to produce final states containing $R$-hadrons. 

The cross-sections are calculated at next-to-leading order in the strong coupling constant (NLO), including the resummation of soft-gluon emission at next-to-leading-logarithmic accuracy (NLO+NLL)~\cite{Beenakker:1996ch,Kulesza:2008jb,Kulesza:2009kq,Beenakker:2009ha,Beenakker:2011fu}. 
The nominal cross-section predictions and the uncertainty are taken from an envelope of cross-section predictions using different PDF sets and factorization and renormalization scales, as described in Ref.~\cite{Kramer:2012bx}.

Selected events containing $R$-hadrons undergo a full detector simulation~\cite{ATL_sim}, where interactions of $R$-hadrons with matter are handled by dedicated \textsc{Geant4}~\cite{Agostinelli:2002hh} routines based on the physics model described in Refs.~\cite{Kraan:2004tz,Mackeprang:2006gx}. This model assumes, for each light valence quark, a cross-section of 12 mb per nucleon and neglects the heavy parton, other than as a reservoir of kinetic energy.
This hadronic scattering is described through a purely phase-space-driven approach.
The probability for a gluino to form a gluon--gluino bound state is assumed to be 10$\%$~\cite{Fairbairn:2006gg}.

The simulation of samples of metastable gluino-based $R$-hadrons is performed similarly to the stable $R$-hadrons. The gluinos within $R$-hadrons are then required to decay via the process 
$\tilde{g}\to\it{q\bar{q}}\tilde{\chi}_1^0$, using \textsc{Pythia 6}.
Gluino masses ($m$) between 400~\GeV\ and 3000~\GeV\ are simulated, with the neutralino mass fixed to 100~\GeV. 
This benchmark decay mode is chosen as it has the highest sensitivity among those studied in Ref.~\cite{SUSY-2014-09}.
Heavier neutralino masses would give lower signal efficiencies, mainly because of the smaller missing transverse momentum available to trigger.
The gluino lifetime ($\tau$) is varied from 0.4~ns to 50~ns.

All simulated samples include a modeling of pileup, adding the expected number of minimum bias proton--proton interactions from the same (in-time pileup) and nearby (out-of-time pileup) bunch crossings. Simulated events are reconstructed using the software used for the collision data.

Additional samples of gluinos are generated for all mass points using \textsc{MG5}{\_a}\textsc{MC@NLO}~\cite{Madgraph}, 
to get a more accurate description of radiative effects than \textsc{Pythia 6} provides. The distribution of the transverse momentum of the gluino--gluino system simulated with \textsc{Pythia 6} is then reweighted in such a way as to match that obtained in samples simulated with \textsc{MG5}{\_a}\textsc{MC@NLO}. This is especially important for stable $R$-hadrons as detailed in Section~\ref{sec:Selection}.


\section{Search Strategy}\label{sec:SearchStategy}
This search is based on the signature of $R$-hadron events containing at least one reconstructed track with large measured $\dEdx$ and high transverse momentum ($p_{\rm T}$). This signature optimizes the search for metastable $R$-hadrons with lifetimes of \emph{O}(ns), as the selection requires that the $R$-hadron only live long enough to traverse the seven layers of the silicon detectors, corresponding to a distance from the beam axis of 37~cm. The search is also sensitive to $R$-hadrons with longer lifetimes, with a small dependence on the modeling of the interaction of $R$-hadrons in the calorimeter and muon systems.

The expected production and interaction of $R$-hadrons in the detector is described in detail in Ref.~\cite{Mackeprang:1385016}. In the models studied in this search and described in Section~\ref{sec:Simulation}, pair-produced gluinos hadronize into two colorless $R$-hadrons with the following charge states: approximately $33\%$ events with two neutral $R$-hadrons, $47\%$ events with one neutral and one charged $R$-hadron, and $20\%$ events with two charged $R$-hadrons. The fragmentation is extremely hard and the $R$-hadrons are therefore isolated. As massive particles with $\beta < 1$, the charged $R$-hadrons are expected to deposit more ionization energy in the detector than typical particles of the same momentum. This ionization signature is used both to select signal events, as described in Section~\ref{sec:Selection}, and to calculate the mass of selected candidates, as described in Section~\ref{sec:Strategy}. The estimated candidate mass is used as the final search discriminant.

If the $R$-hadrons decay before or in the calorimeter, the decay products of $R$-hadrons in models studied here include quarks, which are identified as jets in the calorimeters (see Section~\ref{sec:Selection}), and a 100~\GeV\ neutralino, which carries away significant momentum undetected. If $R$-hadrons traverse the calorimeter without decaying, they are expected to deposit significantly less energy in the calorimeters than SM hadrons of the same momentum, as the heavy parton carries almost all the momentum of the composite particle and limited kinetic energy is available for interactions between the light SM partons and the detector. In the models studied here, over $90\%$ of stable $R$-hadrons, independent of mass, deposit less than $20$~\GeV\ of energy in the calorimeters. However, the electric charge of the $R$-hadron can change due to the hadronic interactions, even if little energy is lost, and the charge of the $R$-hadron leaving the calorimeter is expected to be decoupled from the charge at production~\cite{Mackeprang:1385016}. As hadronic interactions are rare in the tracking detectors, the probability for an $R$-hadron to change its electric charge before the calorimeters is only around 3\%.

Events are selected by a trigger requiring missing transverse momentum ($\MET$), calculated from energy deposits in the calorimeter~\cite{PERF-2011-02}. For events with $R$-hadrons that decay before or inside the calorimeter, undetected neutralinos contribute to the measured transverse momentum imbalance. As stable $R$-hadrons deposit little energy in the calorimeters, most stable $R$-hadron events are selected by the trigger only when QCD initial-state radiation (ISR) boosts the $R$-hadron pair system. Details of the event selection are given in Section~\ref{sec:Selection}.


\section{Data Sample and Event Selection}\label{sec:Selection}
The data sample used for this search was collected while tracking detectors, calorimeters, muon chambers, and magnets were operating normally and corresponds to an integrated luminosity of $\lumi$~fb$^{-1}$ in 2015. The luminosity measurement was calibrated during dedicated beam-separation scans, using the same methodology as that described in Ref.~\cite{DAPR-2011-01}. The uncertainty in the luminosity measurement is $\lumiUnc$. 

Candidate events are selected by an $\MET$ trigger, formed from energy deposits in the calorimeter~\cite{PERF-2011-02}, with a threshold of $70$~GeV. 

The event selection, described in detail below, was optimized relative to the $\sqrt{s}=8$~TeV search~\cite{SUSY-2014-09} to improve the sensitivity to events with $R$-hadrons relative to background events from SM processes. Major improvements include the addition of selections designed to reject tracks from hadrons and electrons, as well as changes to the isolation requirements, which now reject background events from overlapping tracks by identifying clusters in the Pixel or SCT detectors which are consistent with energy deposition from multiple particles. Two signal regions are defined: one targets metastable $R$-hadrons with shorter lifetimes, and the other is optimized for metastable or stable $R$-hadrons that are expected to pass through the muon spectrometer before decaying.

A primary vertex reconstructed from at least two well-reconstructed charged-particle tracks, each with $p_{\rm T} > 400$~\MeV, is required in order to remove noncollision background events. If there is more than one such vertex in an event, the primary vertex is defined as the vertex with the largest scalar sum of associated track momentum. To reduce backgrounds from SM processes, events must have $\MET > 130$~\GeV, where $\MET$ is the offline missing transverse momentum. Jets are reconstructed with the anti-$k_t$ algorithm~\cite{antikt} with radius parameter $R=0.4$ using clusters of energy depositions in the calorimeter as inputs. Events are rejected if they contain a jet with $E_{\rm T} > 20$~\GeV\ that is consistent with noise contributions as determined based on timing and shower shape information. The offline $\MET$ is estimated from reconstructed electrons with $|{\eta}| < 2.47$, reconstructed muons with $|{\eta}| < 2.5$, calibrated jets with $p_{\rm T} > 20$~\GeV\ within $|{\eta}| < 4.5$, and reconstructed tracks which are not associated with other objects and that pass kinematic requirements designed to reject poorly reconstructed tracks.

For signal events, the trigger efficiency varies significantly depending on the model considered. For metastable $R$-hadrons with a mass of 1000~GeV, decaying into a 100~\GeV\ neutralino and quarks, the trigger efficiency increases from $65\%$ for $R$-hadrons with a lifetime of 50 ns to $95\%$ for $R$-hadrons with a lifetime of 0.4 ns. For an $R$-hadron with a lifetime of 10~ns, the trigger efficiency increases slightly with the $R$-hadron mass, from $91\%$ for a mass of 1000~GeV to $96\%$ for a mass of 1600~GeV. For stable $R$-hadrons, the trigger efficiency is approximately $40\%$ for all masses studied. All efficiencies quoted in this section include both the acceptance and the selection
efficiency effects.

The offline $\MET$ calculation includes the momentum of identified muons, which is important to reject background events from SM processes with muons, such as $Z\rightarrow\mu\mu$, that pass the calorimetric $\MET$ trigger. For events with metastable $R$-hadrons, the offline $\MET$ requirement is very efficient (around $95\%$) due to the neutralinos in the event. 
For stable $R$-hadrons that pass the trigger requirement, the offline $\MET$ cut is also efficient (around $90\%$), as the $R$-hadrons do not deposit significant energy in the
calorimeter and in most events the $R$-hadron pair is boosted from initial-state radiation.
Many $R$-hadrons fail to be identified as muons because they are in a neutral charge state going through either the inner detector or the muon spectrometer. Even if a charged $R$-hadron with $\beta < 1$ traverses the muon spectrometer, it can fail to be identified by the standard muon reconstruction algorithm as a prompt muon due to its late time of arrival. In a typical $R$-hadron model considered in this search, only $30\%$ of stable $R$-hadrons with a reconstructed track in the inner detector are identified as prompt muons entering into the $\MET$ calculation.

Selected events must contain at least one candidate $R$-hadron track. For a track to be identified as a candidate, it must have $p_{\rm T} > 50$~\GeV\ and meet the following selection criteria. The track must contain at least seven clusters from silicon detector layers in order to make a good measurement of the track's momentum, must have a cluster in the innermost Pixel layer if expected, and must be associated with the primary vertex. In order to ensure a good ionization measurement, the track must have at least two clusters in the Pixel detector used to measure $\dEdx$. These selection requirements on associated clusters are summarized as cluster requirements.

To reject background events with overlapping tracks that could produce clusters with significant measured ionization, the candidate track must not contain any cluster that is compatible with contributions from two or more tracks. This requirement is over $90\%$ efficient for signal events with lifetimes of 3~ns or greater. For background tracks, it lessens the high ionization tail of the distribution from photon conversions and collimated particles, reducing the size of the background and leaving the remaining tracks with a $\dEdx$ distribution largely independent of background source. In order to reduce backgrounds from particles inside hadronic jets, the scalar sum of the $p_{\rm T}$  ($\sum p_{\rm T}$) of other tracks, with $p_{\rm T} > 1$~\GeV\ and consistent with the primary vertex, in a cone of size $\Delta R < 0.25$,  where $\Delta R \equiv \sqrt{(\Delta \eta)^2 +(\Delta \phi)^2}$, around the candidate track, must be less than 20~\GeV.

The background from SM processes is further reduced by requiring that the candidate track momentum exceeds $150$~\GeV\ and that the relative uncertainty on the momentum measurement be less than $50\%$. The contribution from leptons from $W$ boson decays is reduced by imposing a condition that the transverse mass, $m_{\rm T}$, built with the candidate track and the $\MET$, be greater than $130$~\GeV, where:

\begin{equation}
 m_{\rm T} = \sqrt{2 \pt \MET (1-\cos(\Delta\phi (\MET,\rm{track})) ) }
\end{equation}

and $\Delta\phi (\MET,\rm{track})$ is the azimuthal separation between the track and the $\MET$. This selection has an efficiency between $78\%$ and $89\%$ for $R$-hadrons with lifetimes of 3~ns or greater, and reduces the background by roughly a factor of three. 

Background events due to tracks from electrons and hadrons are reduced by requiring that the energy of any jet associated with the candidate track must be less than the candidate track's measured momentum. Jets are associated with tracks if they are within $\Delta R < 0.05$ of the candidate track and have $p_{\rm T} > 20$~\GeV. Background events containing electrons are further reduced by rejecting candidate tracks associated with a jet with $95\%$ or more of its energy deposited in the electromagnetic calorimeter. These selections are found to be $\>95\%$ efficient for all signal models studied.

Tracks identified as well-reconstructed muons with $p_{\rm T} > 25$~\GeV\ are rejected in the search for metastable $R$-hadrons with lifetimes of $30$~ns or less. $R$-hadrons with a lifetime equal to or greater than $50$~ns typically reach the muon spectrometer, and therefore no such veto is applied in the long-lived $R$-hadron search. To further reject hadronic background in the search for stable $R$-hadrons, the isolation requirement on the $\sum p_{\rm T}$ of nearby tracks is tightened from $20$~\GeV\ to $5$~\GeV\ for candidate $R$-hadron tracks, eliminating $30\%$ of the remaining expected background.

The specific ionization of the candidate track, as measured by the Pixel detector, must be larger than $1.80 - 0.11 |\eta| + 0.17 \eta^2 - 0.05 |\eta|^3 $~\MeV~g$^{-1}$~cm$^{-2}$. This requirement corrects the slight ($5\%$) $|\eta|$ dependence of the $\dEdx$ measurement~\cite{Bichsel88} while maximizing the expected sensitivity to signal events. This selection is approximately $70$--$90\%$ efficient for signal events with lifetimes of 3~ns or greater, depending on the $R$-hadron mass, and retains less than $1\%$ of background events. Additionally, a time-dependent correction is applied to the ionization measurement in data to account for an observed variation in the charge measurement by the IBL electronics due to effects of radiation. The size of this correction is around $5\%$. 

The observed numbers of events in data and the predicted numbers of simulated signal events, for a $1600$~\GeV\ $R$-hadron with a lifetime of $10$~ns, are shown in Table~\ref{tab:cutflowSignal} after each stage of the event selection. The total selection efficiency relative to generated events, which includes both the acceptance and efficiency effects, is also shown for the signal.

\begin{table}[h]
\centering
   \begin{tabular}{| l | r | r |}
          \hline
{\bf Selection level} & {\bf Exp. signal events} & {\bf Observed events in $\lumi$~fb$^{-1}$} \\
    \hline
    \hline
Generated                                                 & 26.0   $\pm$  0.3  &    \\
$\MET$ trigger \& preselection                &  24.8 $\pm$ 0.3  ($95\%$)  &  \\
$\MET > 130$~\GeV\                                  & 23.9 $\pm$  0.3  ($92\%$)   & \\
Track $p_{\rm T} > 50$ and cluster requirements    &   10.7  $\pm$ 0.2 ($41\%$)  &  368324   \\
Isolation requirement &   9.0 $\pm$    0.2  ($35\%$)   & 108079      \\
Track $p > 150$~\GeV\                              &   6.6 $\pm$   0.2  ($25\%$) & 47463     \\
$m_{\rm T} > 130$~\GeV\                                   &   5.8 $\pm$  0.2  ($22\%$)  & 18746    \\
Electron \& hadron veto                          &   5.5$\pm$  0.2 ($21\%$) & 3612    \\
Muon veto                                               & 5.5 $\pm$  0.2 ($21\%$)   & 1668    \\
Ionization requirement                                               & 5.0 $\pm$  0.1 ($19\%$)  &  11     \\
    \hline
    \end{tabular}
    \caption[]{Expected number of events at different steps of the selection for metastable 1600~\GeV\ gluino $R$-hadrons decaying with a 10~ns lifetime, along with the number of events observed in data, for $\lumi$~fb$^{-1}$. The simulated yields are shown with statistical uncertainties only. The total efficiency $\times$ acceptance is also shown in parentheses for the signal.}
    \label{tab:cutflowSignal}
\end{table}


\section{Signal Selection Efficiency}\label{sec:SignalAcceptance}
The overall signal selection efficiency depends on the mass and
lifetime of the signal, ranging from about $19\%$ for a 1600~\GeV\
$R$-hadron with a lifetime of 10 ns to less than $1\%$ for $R$-hadrons
with a lifetime of 0.4 ns. The trigger and offline $\MET$ requirements are more
efficient for $R$-hadrons that decay before or inside the calorimeters. However, as the $R$-hadron lifetime decreases, the probability of
reconstructing the track from the original $R$-hadron in the silicon
detectors decreases. For these two reasons, $R$-hadrons with lifetimes
in the 10--30 ns range have the highest selection efficiency.

The inefficiency for selecting $R$-hadrons with short lifetimes can be
better understood by calculating the selection efficiency relative to
a simple fiducial region that separates out acceptance effects
due to basic kinematic properties of the $R$-hadrons. An event is
considered to be within the fiducial acceptance if it contains at
least one $R$-hadron that passes the following requirements at
MC generator level: charged at production, $p_{\rm T} > 50$~\GeV, $p >
150$~\GeV, $|\eta| < 2.5$, and transverse decay distance $\geq 37$~cm from the
origin. With this definition, the acceptance for $R$-hadron events
with a lifetime of 1~ns ranges from $7\%$ for a mass of 600~\GeV\ to
$4\%$ for a mass of 1600~\GeV. The selection efficiency for $R$-hadron events in the fiducial region
has only a mild dependence on the mass; in the $R$-hadron mass range
from 600 to 1600 GeV, the selection efficiences are $20\%$ to $25\%$
($\tau = 1$~ns),  $35\%$ to $39\%$ ($\tau = 3$~ns),  $39\%$ to $45\%$
($\tau = 10$~ns),  and $31\%$ to $35\%$ ($\tau = 30$~ns), respectively.

Table~\ref{tab:efficiencySyst} summarizes the systematic uncertainties
in the predicted signal yields. The systematic
uncertainties are calculated and used independently for each mass
bin and signal lifetime. In the table, only the maximum uncertainty is quoted.

The trigger and offline $\MET$ selections are sensitive to the
modeling of ISR in the event, in particular for signal events that do
not decay inside the detector volume. Half of the
difference between the selection efficiency of the \textsc{Pythia 6} samples
and those reweighted with \textsc{MG5}{\_a}\textsc{MC@NLO} is taken as an uncertainty due
to radiative effects.
 
The $\MET$ trigger modeling is studied in a sample of
$Z\rightarrow\mu\mu$ events and an uncertainty is
assigned based on differences between efficiencies measured in data
and in simulation. The uncertainty in the $\MET$ calibration comes
primarily from uncertainties on the jet energy scale. A systematic uncertainty in the
measurement of the Pixel ionization is evaluated by comparing the
ionization of an inclusive sample of simulated and observed low-momentum tracks. An asymmetric uncertainty
is assigned due to the selection efficiency difference observed after the most probable value of
the signal distribution is reweighted based on the comparison of low-momentum tracks between data and simulation. The
uncertainty in the signal acceptance is asymmetric since the
ionization in simulation is systematically larger than that observed
in data. An uncertainty in the efficiency of the muon veto is estimated
by studying the modeling of the muon spectrometer's timing
measurement.

The signal selection efficiency is found to be largely independent of the number of pileup
events in simulation. However, to cover any remaining
difference, efficiency differences from variations in the pileup distributions
are accounted for as an additional source of uncertainty. The uncertainty
on the momentum scale and resolution of the tracks results in a small
uncertainty in the selection efficiency. Variations of the requirements to
reject tracks from electrons and hadrons are found to have a negligible effect
on the signal efficiency. The theoretical uncertainty in the signal
cross-section is described in Section~\ref{sec:Simulation}.

\begin{table}[!htbp]
\centering
    \begin{tabular}{|l|cc|}
       \hline
    {\bf Source of uncertainty} & $-$[\%]& $+$[\%]\\
    \hline
    \hline
   ISR modeling ($R$-hadron stable) & 14 & 14 \\
   ISR modeling  ($R$-hadron metastable) & 1.5 & 1.5  \\
   Trigger turn-on & 0.9 & 0.9 \\ 
    $\MET$ scale & 1.1 & 2.2\\
    Pileup  & 1.1 & 1.1  \\
    Ionization parameterization & 7.1 & 0 \\  
    Momentum parameterization  &0.3 & 0.3 \\
    $\mu$~identification (metastable only)  & 4.3 & 4.3 \\ 
        \hline
    {\bf Total systematic uncertainty in acceptance $\times$ efficiency}&& \\
     \emph{\textbf{Stable $R$-hadron}} & 16 & 14 \\  
    \emph{\textbf{Metastable $R$-hadron}} & 9 & 5 \\  
    \hline
    Luminosity&  $5$&5\\ 
    Signal cross-section & 28 & 28 \\
       \hline
    \end{tabular}
    \caption[]{Systematic uncertainties on the predicted signal yields. The uncertainty depends on the mass and on the
      lifetime; only the maximum negative and positive values across
      all signal models (see Section~\ref{sec:Simulation}) are reported here. \label{tab:efficiencySyst}}
\end{table}


\section{Background Estimation}\label{sec:Background}
The expected shape and normalization of the mass distribution of background events from SM processes is derived from data. The distributions of key variables are extracted in two control regions in data and these templates are used to generate the expected background distribution in the signal region. The choice of control samples takes into account the measured correlations between the key variables $p$, $\dEdx$ and $\eta$, and the control region selections minimize the possible signal contamination. The same regions are used in the searches for stable and metastable particles, except for the rejection of track candidates identified as muons in the latter case and for the tighter track isolation requirement in the former case.
   
The first control region (hereafter called CR1) is selected by applying all the selections described in Section~\ref{sec:Selection}, except for the requirement on high ionization, which is inverted. The tracks in this control region are kinematically similar to those in the signal region for SM processes, and the expected $p$ and $\eta(p)$ distributions of tracks from background processes are extracted from this region. The second sample (CR2 hereafter) is selected by inverting the $\MET$ requirement ($\MET<130$~\GeV), while keeping all other selections unchanged. As more than $90\%$ of $R$-hadron events that pass the trigger have $\MET>130$~\GeV, while the $\MET$ distribution from SM background processes falls steeply in the range $70$--$130$~\GeV, CR2 is background-dominated. Tracks from this region are used to derive $\dEdx$ templates in bins of $\eta$, as $\MET$ and $\dEdx$ are uncorrelated for tracks from SM processes. The signal contamination is less than $1\%$ in both control regions for all $R$-hadron masses and lifetimes considered in this search.

The shape of the background in the signal region is estimated from the control region templates with the following procedure: 1 million $(p,\ \eta, \ \dEdx)$ triplets 
are generated, where the momentum is randomly sampled according to the template from CR1 tracks, 
the pseudorapidity is generated according to the $\eta(p)$-binned templates based on CR1 tracks, and 
the ionization is sampled according to $\dEdx(\eta)$-binned templates from CR2 tracks. This procedure maintains the measured correlations between $p$, $\eta$, and $\dEdx$ for tracks from SM sources. The particle mass $M$ is calculated given the predicted $\dEdx$ and $p$ values, using the technique explained in Section~\ref{sec:Strategy}.

\begin{figure}[htbp]
\centering
\resizebox{0.65\textwidth}{!}{%
  \includegraphics{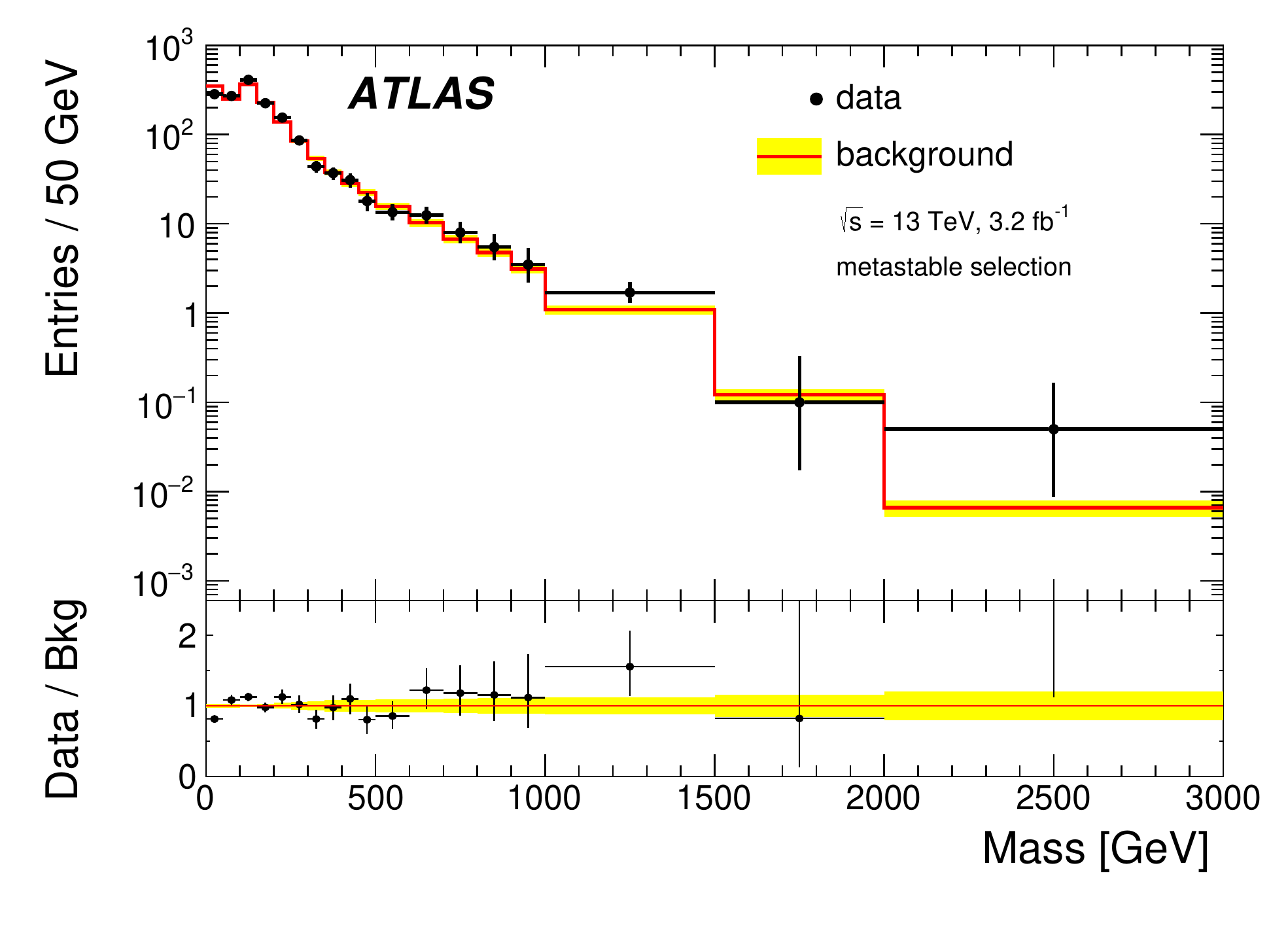}  
  }
  \caption{Expected mass distribution from background events in the signal region for the metastable $R$-hadrons, before the ionization requirement is imposed. Data are superimposed. The yellow band around the background estimation includes both the statistical and systematic uncertainties. The ratio of data to expected background is shown in the lower panel with statistical uncertainties only.}
  \label{fig:metastable_before_ion}
\end{figure}

The normalization of the generated background is obtained by scaling the background to the data
in the shoulder region of the mass distribution (i.e. $M<$~160~\GeV), where a possible signal has already been excluded~\cite{SUSY-2014-09}. The normalization is performed on the samples before the ionization requirement. The expected mass distributions of the background and the observed data, before the ionization requirement, are shown in Figure~\ref{fig:metastable_before_ion} for the metastable $R$-hadrons selection.

Systematic uncertainties in the background estimate are evaluated by varying the relative fraction of muons in the control regions, using an analytical description to vary the shape of the high ionization tail in the $\dEdx$ template, and by changing the IBL ionization correction by $\pm 1 \sigma$. The statistical uncertainty in the template shape, which is estimated by Poisson fluctuations of the templates, dominates. There is also a small statistical uncertainty from the size of the normalization region. The breakdown of uncertainties in the background estimate is shown in Table~\ref{tab:BkgSysSummary}.  

\begin{table}
\centering
    \begin{tabular}{|l|r|}
       \hline
    {\bf Source of uncertainty} &  [\%]\\
    \hline
    \hline
   Statistical uncertainty from control region templates & 15 \\
   Statistical uncertainty from normalization region & 3 \\
   Analytical description of $\dEdx$ & 4  \\	
   Particle species composition of CR2 & 3 \\
   IBL ionization correction & 4 \\
    \hline
    \end{tabular}
        \caption[]{Summary table of the uncertainties in the background estimate. The uncertainties are evaluated integrating over the mass windows considered for the signal search (see Section~\ref{sec:results}). If the uncertainty depends on the mass, the largest value is quoted.
        \label{tab:BkgSysSummary}}
\end{table}

The complete procedure is tested on signal-depleted validation regions. The validation regions are selected with the same requirements as for the nominal signal and control regions, but requiring the track momentum to be in the range $50<p<150$~\GeV. Applied to these validation samples, the procedure described above yields a calculated background of 20.0 $\pm$ 3.0 (stat.) $\pm$ 1.2 (syst.) events in the validation region for metastable $R$-hadrons, while 14 events are observed in the data.  For the control region for stable $R$-hadrons, 28.1 $\pm$ 4.2 (stat.) $\pm$ 1.9 (syst.) events are estimated from this procedure and 20 are observed. While the observed numbers of events in the validation regions are lower than predicted, they are consistent with statistical fluctuations at the level of $1.7\sigma$.


\section{Results}\label{sec:results}
\begin{figure}[htbp]
\resizebox{0.65\textwidth}{!}{%
 	  \includegraphics{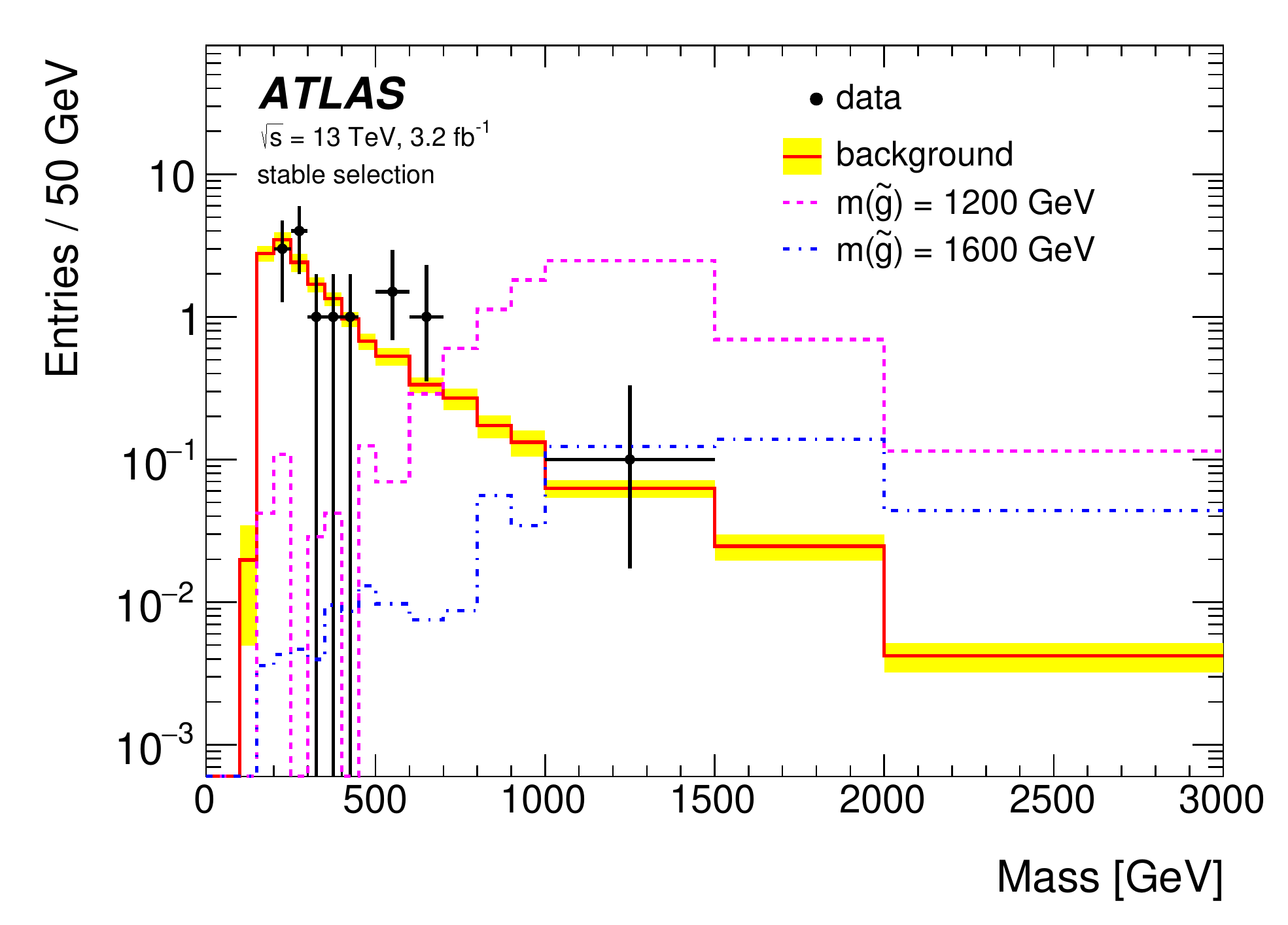}
}
\resizebox{0.65\textwidth}{!}{%
	  \includegraphics{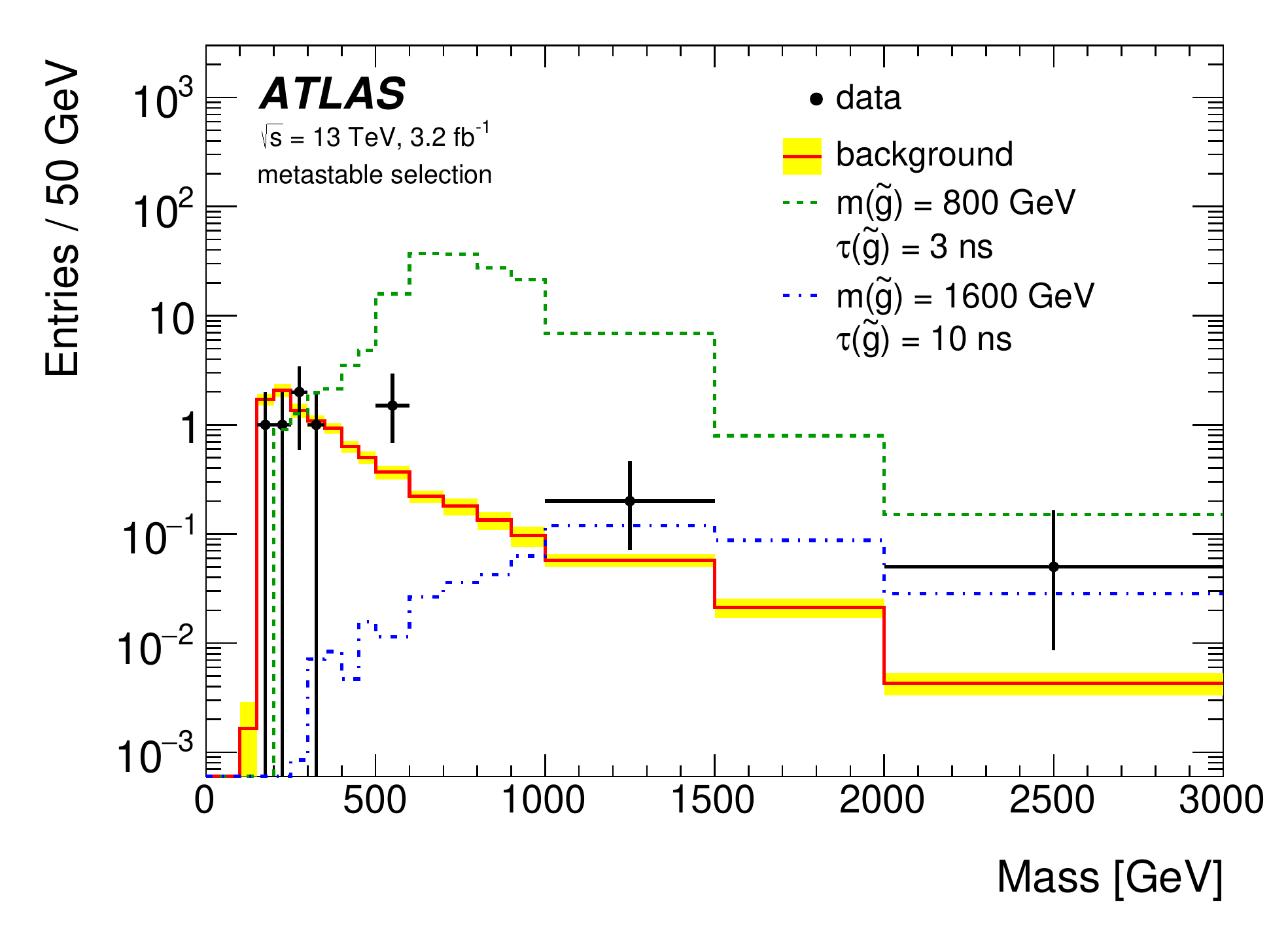}\\
}
\centering
\caption{Mass distribution for data and background for stable (top)
  and metastable (bottom) particle searches. The yellow band around the
  background estimation includes both the statistical and systematic
  uncertainties. Also shown are two examples for signals as expected for gluino
$R$-hadrons in the explored mass range.}
\label{fig:finalMass}
\end{figure}

In data, 16 events are observed for the stable $R$-hadron selection
and 11 are observed for the metastable
selection. Table~\ref{tab:bkgSummary} summarizes the background
estimates with total statistical and systematic uncertainty as well as the observed events for the 
metastable and stable $R$-hadron selection. The observed mass distribution in data
is shown in Figure~\ref{fig:finalMass} for both selection regions,
along with the background expectation. 

\begin{table}[htbp]
\centering
   \begin{tabular}{| l | c | c |}
       \hline
{\bf Selection region} & {\bf Background exp.} & {\bf Data} \\
\hline
\hline
Metastable $R$-hadron     & 11.1  $\pm$ 1.7 $\pm$ 0.7  &  11    \\
Stable $R$-hadron     &  17.2 $\pm$ 2.6 $\pm$ 1.2   &  16   \\
    \hline
    \end{tabular}
    \caption[]{Estimated number of background events and the number of observed events in data in the final selection regions. The background predictions show both the statistical and systematic uncertainties.
    \label{tab:bkgSummary}}
\end{table}

No evidence of a signal above the background is observed. 
Expected and observed upper limits on $R$-hadron production cross-sections
are evaluated at 95\% CL for a discrete set of $R$-hadron
mass values by counting simulated signal, background and observed data
events that survive the selection in a $\pm1.4~\sigma$ wide mass window
around the position of each simulated signal mass peak for the
corresponding mass hypothesis being probed. The peak position and width
are estimated by a Gaussian fit to each individual signal
mass distribution ($\sigma\approx$280~[$\approx$500]~\GeV\ at an
$R$-hadron mass of 1200~[1600]~\GeV). When the lifetime of the $R$-hadron is long enough to reach the muon
spectrometer, the stable selection becomes more efficient than the
metastable one and improves the expected sensitivity. This transition
happens around $50$~ns. The choice of whether to apply the metastable or
stable $R$-hadron selection to each lifetime is based on the best
expected limit. Based on this procedure, the stable selection is applied to both the $50$~ns
and stable lifetimes.

The cross-section upper limits are extracted, separately for the stable and the metastable searches (in the latter case,
for each lifetime value), with the $CL_{S}$ method~\cite{Read:2002hq},
using the profile likelihood ratio as a test statistic.
In the procedure, the systematic uncertainties in
the signal and background yields, as evaluated in Section~\ref{sec:SignalAcceptance} and Section~\ref{sec:Background},
are treated as Gaussian-distributed nuisance parameters. The
statistical uncertainty in the background distribution also takes into account
the uncertainty due to the normalization. Lower limits on the $R$-hadron mass are
then derived, for each lifetime, by comparing the
measured cross-section upper limits to the theoretically predicted
production cross-section.
The resulting lower limits set on the mass of the stable and metastable particles are summarized in Table~\ref{tab:massexcl}.
\begin{table}[htp]
\centering
   \begin{tabular}{|cccc| }
       \hline
 {\bf Selection}& {\bf $\tau$ [ns] }& {\bf $M_{\mathrm{obs}}>$[\GeV] }& {\bf $M_{\mathrm{exp}}>$[\GeV]}\\
\hline
   \hline
Metastable   & 0.4       & 740       & 730 \\
        "   & 1.0       & 1110       & 1150 \\
        "   & 3.0       & 1430       & 1470\\
        "   & 10        & 1570       & 1600 \\
        "   & 30         & 1580      & 1620 \\
        "  & 50         & 1540      & 1590 \\
 \hline
Stable   & 50         & 1590      & 1590 \\
     "        & stable           & 1570     & 1580 \\
\hline
  \end{tabular}
\caption{Observed and expected 95\% CL lower limit on mass is listed for gluino~$R$-hadrons that are either stable or decaying to $q\bar{q}$ plus a light neutralino of mass $m(\tilde{\chi}_0)=100$~\GeV. The stable selection is more efficient for lifetimes of 50~ns or more.}
   \label{tab:massexcl}
\end{table}%

\begin{figure}[htbp]
\centering
\resizebox{0.65\textwidth}{!}{%
  \includegraphics{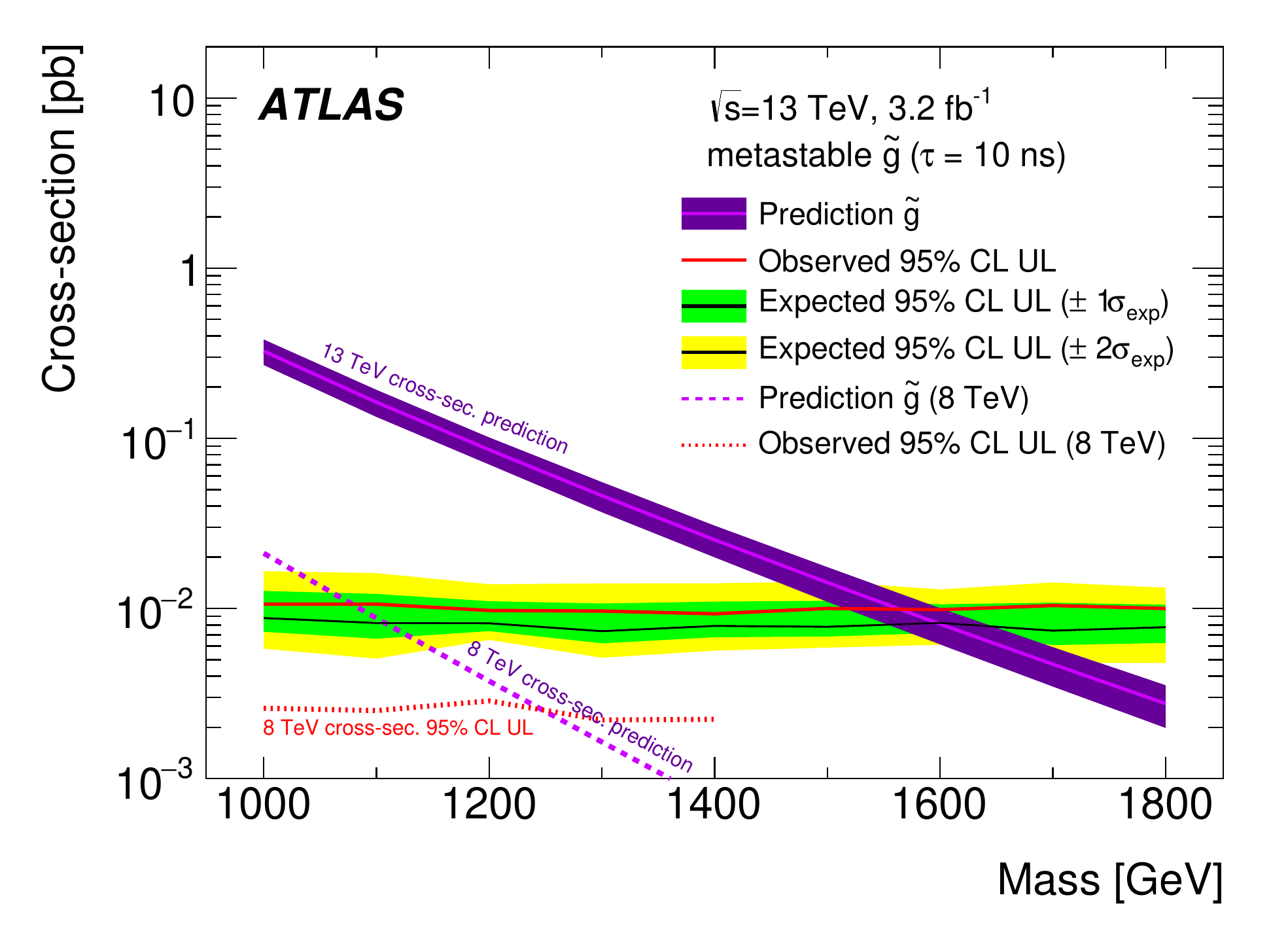}  
  }
  \caption{
  cross-section as a function of mass for gluino $R$-hadrons with lifetime $\tau$=10~ns, decaying to $q\bar{q}$ plus a light neutralino of mass $m(\tilde{\chi}_0)=100$~\GeV.  Theoretical values for the cross-section are shown with their error. The expected upper limits (UL) in the case of background only is shown by the solid black line, with its $\pm 1 \sigma$ and $\pm 2 \sigma$ bands, green and yellow respectively. The observed 95\% UL is shown as a solid red line. The cross-section for the same process and the corresponding  95\% UL measured at $\sqrt{s} = 8$~\TeV~\cite{SUSY-2014-09} are shown as dashed lines.}
  \label{fig:limmass_gluino10ns}
\end{figure}

The strongest mass limits are obtained for lifetimes of 10~ns or more. The cross-section upper limits for 10~ns lifetime $R$-hadrons decaying
to $q\bar{q}$ plus a light neutralino of mass
$m(\tilde{\chi}_0)$=100~\GeV\ are shown in
Figure~\ref{fig:limmass_gluino10ns}. For this signal, the expected and
observed lower mass limits are 1600~\GeV\ and 1570~\GeV, respectively. 
The 8~\TeV\ result excluded $R$-hadrons with a 10~ns lifetime and masses
up to 1185~\GeV, using the lower edge of the $\pm 1\sigma$ band around
the theoretical cross-section. 

\begin{figure}[htbp]
\centering
\resizebox{0.65\textwidth}{!}{%
  \includegraphics{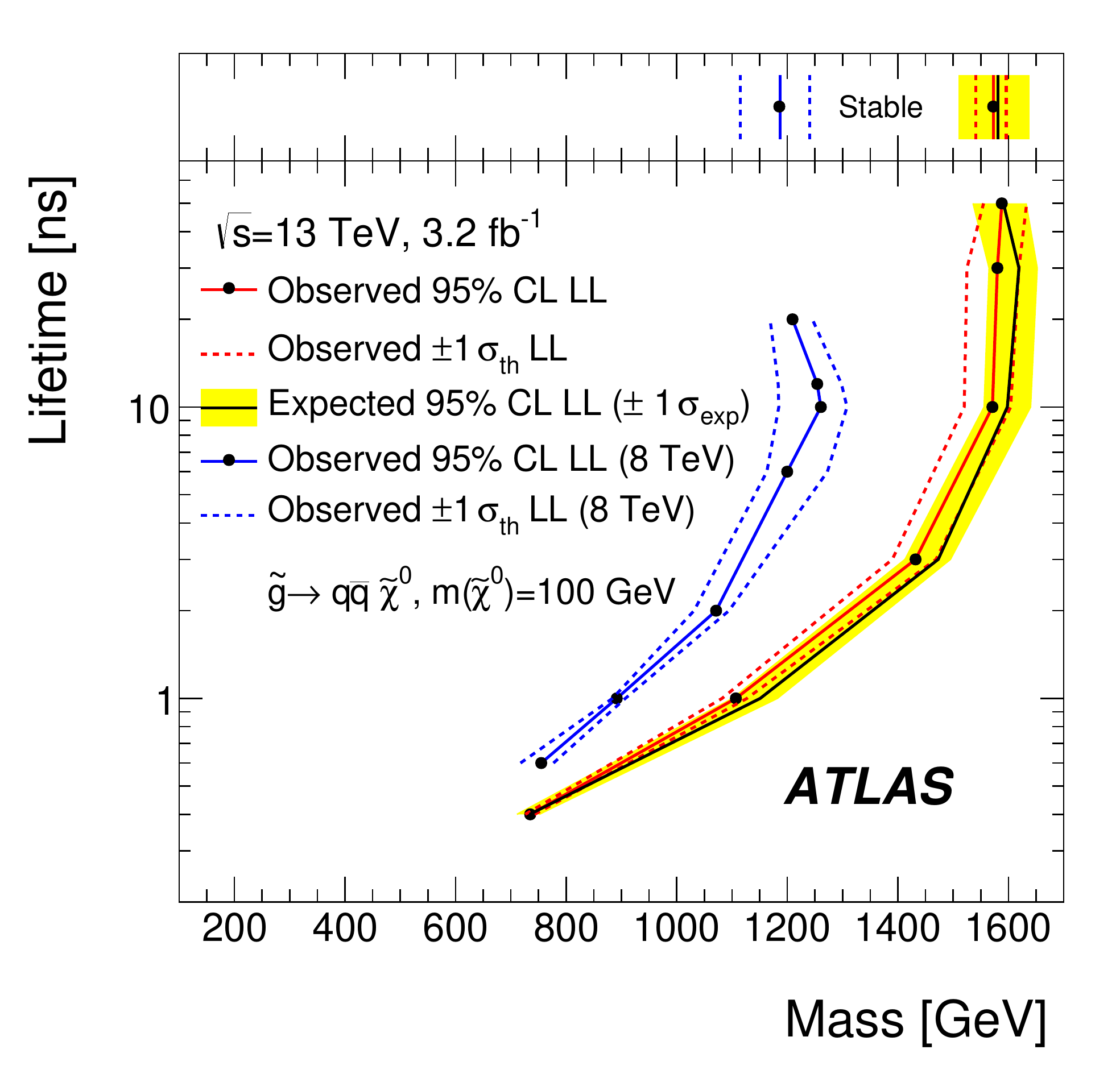}  
  } 
\caption{Excluded range of lifetimes as a function of gluino $R$-hadron mass. 
The expected lower limit (LL), with its experimental $\pm 1 \sigma$ band, is given with respect
to the nominal theoretical cross-section. The observed 95\% LL obtained at $\sqrt{s} = 8$~\TeV~\cite{SUSY-2014-09} 
 is also shown for comparison.}
\label{fig:LifetimeVsMassExclusionPlot}
\end{figure}

The excluded regions on the lifetime--mass plane for $R$-hadrons decaying
to $q\bar{q}$ plus a light neutralino of mass $m(\tilde{\chi}_0)$=100~\GeV\ 
are shown in Figure~\ref{fig:LifetimeVsMassExclusionPlot}.
 
\section{Conclusion}
\label{sec:conclusion}
This paper presents the results of a search for massive, long-lived particles with lifetimes of \emph{O}(ns) or longer, produced in \lumi~fb$^{-1}$ of $pp$ collision data at $\sqrt{s} = 13$~\TeV\ at the LHC, and identified by anomalous specific ionization energy loss in the ATLAS Pixel detector. This analysis takes advantage of the addition of a fourth silicon layer to the Pixel detector, the increased parton luminosity, and analysis improvements to significantly extend a search performed at $\sqrt{s} = 8$~\TeV~\cite{SUSY-2014-09}. No excess of events is observed over the background estimate. 
Gluino $R$-hadrons with lifetimes above 0.4~ns and decaying to $q\bar{q}$ plus a 100~\GeV\ neutralino are excluded at the $95\%$ 
confidence level with lower mass limit range between 740~\GeV\ and 1590~\GeV. 
In the case of stable $R$-hadrons the lower mass limit at the $95\%$ confidence level is 1570~\GeV.
The observed lower limit on $R$-hadron masses increases by up to approximately 400~\GeV\ relative to the equivalent analysis at $\sqrt{s} = 8$~\TeV. 

\section*{Acknowledgments}

We thank CERN for the very successful operation of the LHC, as well as the
support staff from our institutions without whom ATLAS could not be
operated efficiently.

We acknowledge the support of ANPCyT, Argentina; YerPhI, Armenia; ARC, Australia; BMWFW and FWF, Austria; ANAS, Azerbaijan; SSTC, Belarus; CNPq and FAPESP, Brazil; NSERC, NRC and CFI, Canada; CERN; CONICYT, Chile; CAS, MOST and NSFC, China; COLCIENCIAS, Colombia; MSMT CR, MPO CR and VSC CR, Czech Republic; DNRF and DNSRC, Denmark; IN2P3-CNRS, CEA-DSM/IRFU, France; GNSF, Georgia; BMBF, HGF, and MPG, Germany; GSRT, Greece; RGC, Hong Kong SAR, China; ISF, I-CORE and Benoziyo Center, Israel; INFN, Italy; MEXT and JSPS, Japan; CNRST, Morocco; FOM and NWO, Netherlands; RCN, Norway; MNiSW and NCN, Poland; FCT, Portugal; MNE/IFA, Romania; MES of Russia and NRC KI, Russian Federation; JINR; MESTD, Serbia; MSSR, Slovakia; ARRS and MIZ\v{S}, Slovenia; DST/NRF, South Africa; MINECO, Spain; SRC and Wallenberg Foundation, Sweden; SERI, SNSF and Cantons of Bern and Geneva, Switzerland; MOST, Taiwan; TAEK, Turkey; STFC, United Kingdom; DOE and NSF, United States of America. In addition, individual groups and members have received support from BCKDF, the Canada Council, CANARIE, CRC, Compute Canada, FQRNT, and the Ontario Innovation Trust, Canada; EPLANET, ERC, FP7, Horizon 2020 and Marie Sk{\l}odowska-Curie Actions, European Union; Investissements d'Avenir Labex and Idex, ANR, R{\'e}gion Auvergne and Fondation Partager le Savoir, France; DFG and AvH Foundation, Germany; Herakleitos, Thales and Aristeia programmes co-financed by EU-ESF and the Greek NSRF; BSF, GIF and Minerva, Israel; BRF, Norway; the Royal Society and Leverhulme Trust, United Kingdom.

The crucial computing support from all WLCG partners is acknowledged
gratefully, in particular from CERN and the ATLAS Tier-1 facilities at
TRIUMF (Canada), NDGF (Denmark, Norway, Sweden), CC-IN2P3 (France),
KIT/GridKA (Germany), INFN-CNAF (Italy), NL-T1 (Netherlands), PIC (Spain),
ASGC (Taiwan), RAL (UK) and BNL (USA) and in the Tier-2 facilities
worldwide.

\printbibliography

\newpage 
\input{atlas_authlist}

\end{document}

%% file: atlas_authlist.tex
\begin{flushleft}
{\Large The ATLAS Collaboration}

\bigskip

M.~Aaboud$^\textrm{\scriptsize 136d}$,
G.~Aad$^\textrm{\scriptsize 87}$,
B.~Abbott$^\textrm{\scriptsize 114}$,
J.~Abdallah$^\textrm{\scriptsize 65}$,
O.~Abdinov$^\textrm{\scriptsize 12}$,
B.~Abeloos$^\textrm{\scriptsize 118}$,
R.~Aben$^\textrm{\scriptsize 108}$,
O.S.~AbouZeid$^\textrm{\scriptsize 138}$,
N.L.~Abraham$^\textrm{\scriptsize 150}$,
H.~Abramowicz$^\textrm{\scriptsize 154}$
% ...
\bigskip
\\
% ...
$^{12}$ Institute of Physics, Azerbaijan Academy of Sciences, Baku, Azerbaijan\\
% ...
$^{65}$ University of Iowa, Iowa City IA, United States of America\\
% ...
$^{87}$ CPPM, Aix-Marseille Universit{\'e} and CNRS/IN2P3, Marseille, France\\
% ...
$^{108}$ Nikhef National Institute for Subatomic Physics and University of Amsterdam, Amsterdam, Netherlands\\
% ...
$^{114}$ Homer L. Dodge Department of Physics and Astronomy, University of Oklahoma, Norman OK, United States of America\\
% ...
$^{118}$ LAL, Univ. Paris-Sud, CNRS/IN2P3, Universit{\'e} Paris-Saclay, Orsay, France\\
% ...
$^{136}$ $^{(a)}$ Facult{\'e} des Sciences Ain Chock, R{\'e}seau Universitaire de Physique des Hautes Energies - Universit{\'e} Hassan II, Casablanca; $^{(b)}$ Centre National de l'Energie des Sciences Techniques Nucleaires, Rabat; $^{(c)}$ Facult{\'e} des Sciences Semlalia, Universit{\'e} Cadi Ayyad, LPHEA-Marrakech; $^{(d)}$ Facult{\'e} des Sciences, Universit{\'e} Mohamed Premier and LPTPM, Oujda; $^{(e)}$ Facult{\'e} des sciences, Universit{\'e} Mohammed V, Rabat, Morocco\\
% ...
$^{138}$ Santa Cruz Institute for Particle Physics, University of California Santa Cruz, Santa Cruz CA, United States of America\\
% ...
$^{150}$ Department of Physics and Astronomy, University of Sussex, Brighton, United Kingdom\\
% ...
$^{154}$ Raymond and Beverly Sackler School of Physics and Astronomy, Tel Aviv University, Tel Aviv, Israel\\
% ...
$^{a}$ Also at Department of Physics, King's College London, London, United Kingdom\\
$^{b}$ Also at Institute of Physics, Azerbaijan Academy of Sciences, Baku, Azerbaijan\\
$^{c}$ Also at Novosibirsk State University, Novosibirsk, Russia\\
$^{d}$ Also at TRIUMF, Vancouver BC, Canada\\
$^{e}$ Also at Department of Physics {\&} Astronomy, University of Louisville, Louisville, KY, United States of America\\
% ...
\end{flushleft}
